\documentclass[a4paper]{jpconf}

\usepackage{xspace} % clever space in aliases
\usepackage{graphicx}
\usepackage{amsmath}
\usepackage{caption}
\usepackage{subcaption}
\usepackage{siunitx}
\DeclareSIUnit\permille{\text{\textperthousand}}

\graphicspath{{figures/}}

\newcommand*{\apx}{\emph{ATLASpix\_Simple}\xspace}

\usepackage{lineno}
\begin{document}
\setlength{\abovedisplayskip}{4pt}
\setlength{\belowdisplayskip}{4pt}
\setlength{\abovedisplayshortskip}{0pt}
\setlength{\belowdisplayshortskip}{4pt}
\setlength{\textfloatsep}{10pt plus 2pt minus 2pt}

\title{Performance Studies of the ATLASpix HV-MAPS Prototype for Different Substrate Resistivities}

\author{Jens Kr\"oger$^{1,2}$ on behalf of the CLICdp Collaboration}

\address{$^1$Physikalisches Institut der Universit\"at Heidelberg, Im Neuenheimer Feld 226, 69120 Heidelberg, Germany}
\address{$^2$EP-DT-TP, CERN, Esplanade des Particules 1, 1211 Meyrin, Switzerland}

\ead{kroeger@physi.uni-heidelberg.de}

\begin{abstract}
The \emph{ATLASpix} high-voltage monolithic active pixel sensor (HV-MAPS) was designed as a technology demonstrator for the ATLAS ITk Upgrade and the CLIC tracking detector.
In this contribution new results from laboratory-based energy calibration measurements using fluorescence X-rays are presented for the \apx matrix.
These findings are complemented by new results from test-beam studies with inclined tracks, in which the active charge collection depth is determined.
\end{abstract}

\section{Introduction}
The experimental conditions at future high-energy particle physics experiments, such as the High-Luminosity Large Hadron Collider (HL-LHC)~\cite{hl-lhc-tdr} or the Compact Linear Collider (CLIC)~\cite{clic-report}, require highly performant detector systems to meet the foreseen physics goals.
Due to the advances in the silicon sensor industry, all-silicon detector systems are regarded attractive options for future tracking detectors.
Monolithic technologies combine both the sensor and the readout electronics on one chip, resulting in a reduced material budget compared to hybrid technologies.
They are considered particularly suitable for large-area applications due to their cost efficiency and large-scale production capabilities of the CMOS imaging industry.

The \emph{ATLASpix}~\cite{peric-apx} was designed as a technology demonstrator for the ATLAS ITk upgrade~\cite{atlas-itk} and the CLIC tracking detector~\cite{yellowreport_detectortech}.
It was manufactured in a commercial \SI{180}{nm} HV-CMOS process on wafers with different substrate resitivities ranging from \SI{20}{\ohm cm} to \SI{200}{\ohm cm}.
As a high-voltage monolithic active pixel sensor (HV-MAPS), it features a fully integrated readout.
A high bias voltage of $\mathcal{O}(\SI{100}{V})$ leads to large signals due to a large depleted volume, as well as a high electric field resulting in fast charge collection via drift.
By removing bulk material from the backside, the sensors can be thinned to \SI{50}{\micro m}.
The active matrix of the \apx consists of 25 columns and 400 rows of pixel cells with a pitch of $130\times\SI{40}{\micro m^2}$, which are read out in a zero-suppressed triggerless column drain scheme.
Each pixel consists of a deep N-well in a p-substrate forming the sensor diode.
The N-well contains the in-pixel electronics comprising a charge-sensitive amplifier and a comparator.
For each hit, the time-of-arrival (ToA) with a resolution of \SI{10}{bits} and a binning of \SI{16}{ns}, as well as the time-over-threshold (ToT) with a 6-bit resolution are recorded.

\section{Energy Calibration with Fluorescence X-Rays}
In order to perform an energy calibration of the detection threshold and the time-over-threshold (ToT) measurement, the \apx samples were exposed to X-rays with well-defined energies.
A commercial X-ray tube was used to excite fluorescence a target placed in front of the X-ray tube.
By choosing different target materials (titanium, iron, copper), sharp $K_\alpha$ peaks with well-defined energies were generated to which the \apx were exposed.

\subsection{Analysis Method}
\begin{figure}[b]
\begin{center}
\begin{subfigure}[b]{0.3\textwidth}
\includegraphics[width=\textwidth]{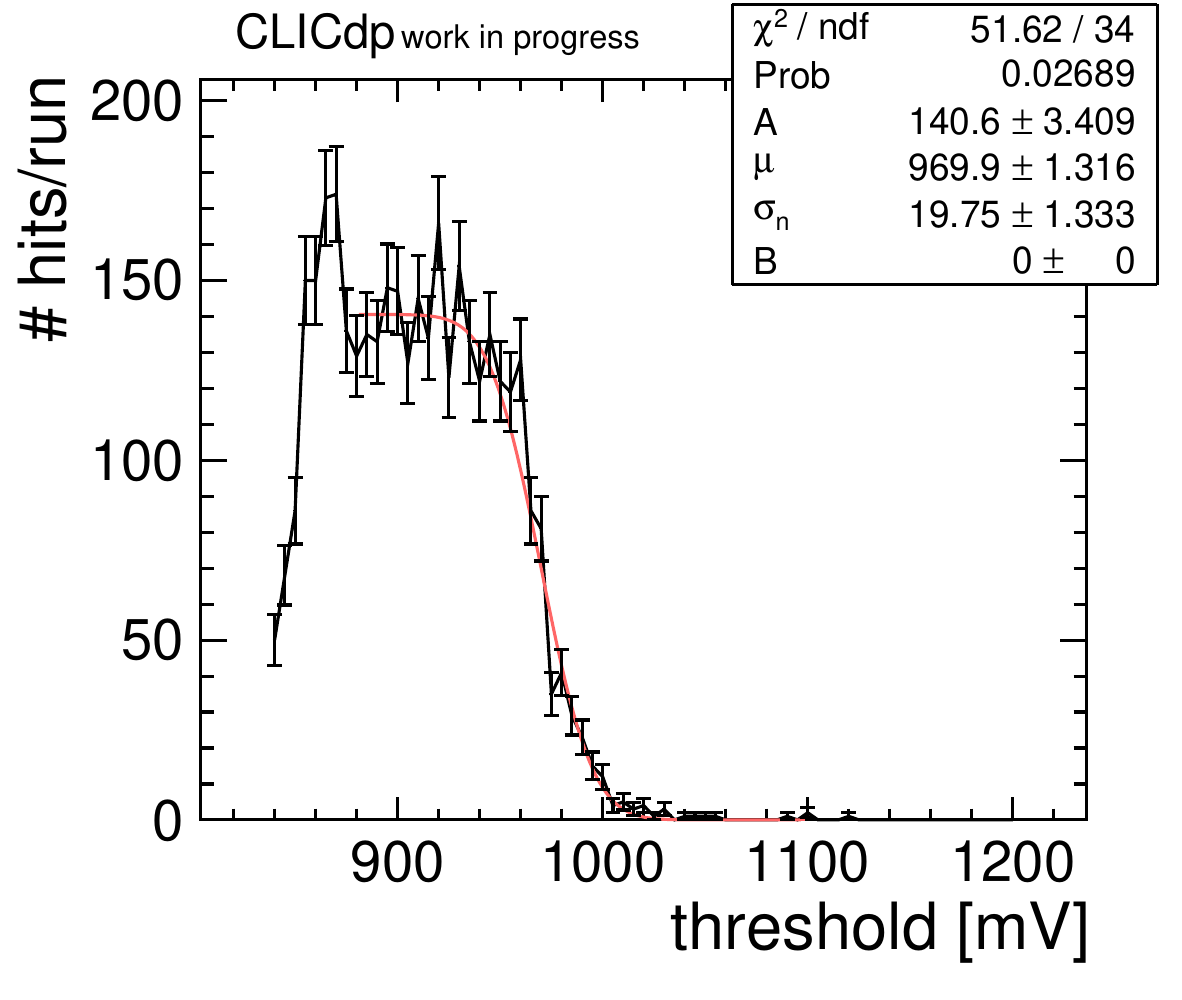}
\caption{\label{fig:example-scurve}Exemplary s-curve fit for pixel (10,10).\hfill\newline}
\end{subfigure}\hfill%
\begin{subfigure}[b]{0.3\textwidth}
\includegraphics[width=\textwidth]{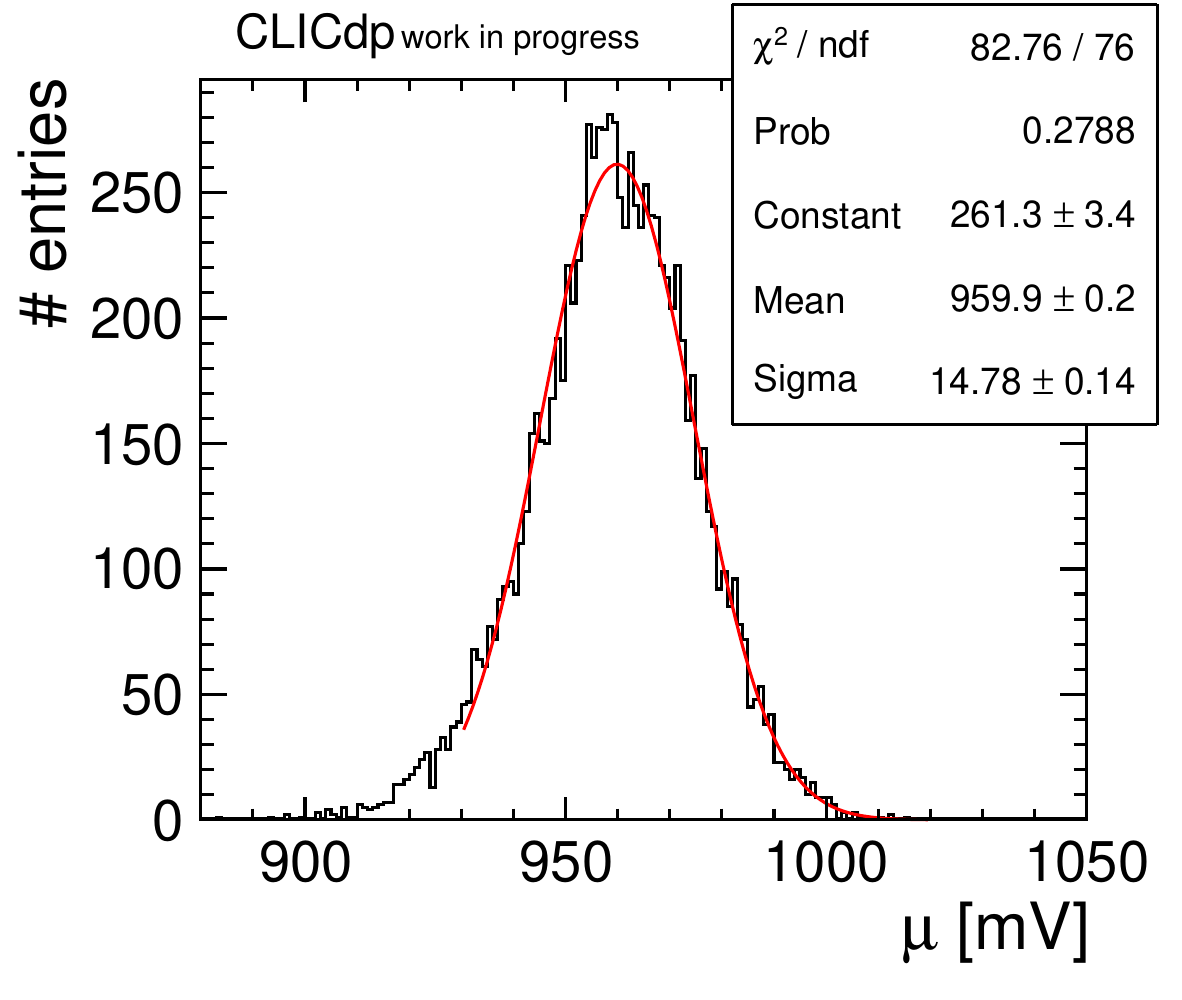}
\caption{\label{fig:mu1D}Distribution of the threshold $\mu$ for all pixels fitted with a Gaussian.}
\end{subfigure}\hfill%
\begin{subfigure}[b]{0.3\textwidth}
\includegraphics[width=\textwidth]{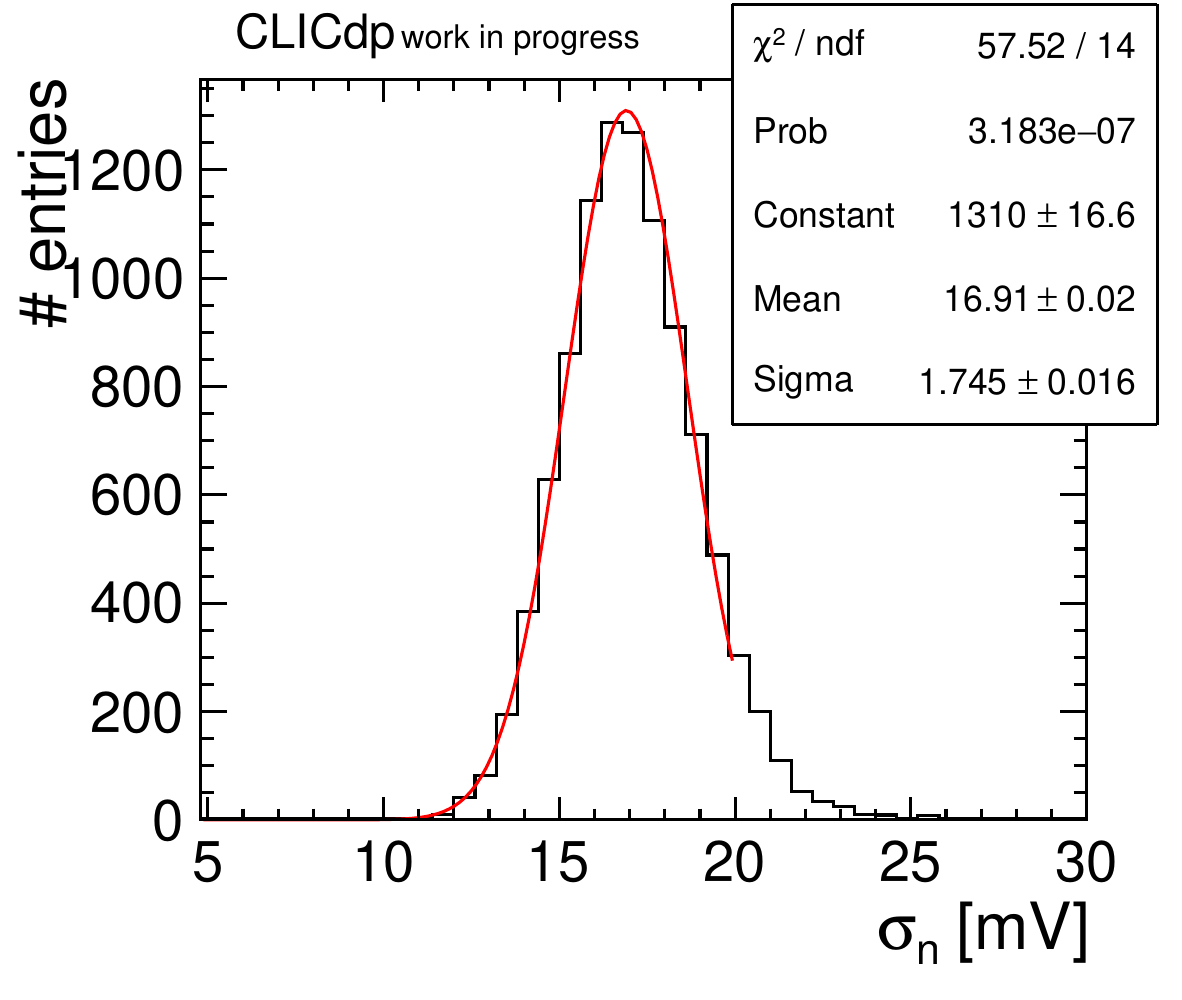}
\caption{\label{fig:sigma1D}Distribution of the noise $\sigma_n$ for all pixels fitted with a Gaussian.}
\end{subfigure}
\vspace{-0.25cm}
\caption{Fit results obtained from s-curve fits for each pixel for a \SI{200}{\ohm cm} sample with fluorescent X-rays from an iron target (\SI{6.4}{keV}) at a bias voltage of \SI{-50}{V}.}
%\vspace{-0.5cm}
\end{center}
\end{figure}

For each pixel of the matrix, the number of pixel hits per run of \SIrange{20}{40}{s} is plotted against the threshold.
This yields a distribution as shown in Figure~\ref{fig:example-scurve}, which can be described by a so-called s-curve:
\begin{linenomath*}
\begin{equation}
\label{eq:s-curve}
f_s(x) = \frac{A}{2} \left( 1 - \text{erf}\left( - \frac{x- \mu}{\sqrt{2}\sigma_n} \right) \right) + B \text{, with the error function: } \text{erf}(x) = \frac{2}{\sqrt{\pi}} \int_{0}^{x} e^{-t^2} \mathrm{d}t,
\end{equation}
\end{linenomath*}
where $A$ is a normalisation constant, $\mu$ is the threshold value corresponding to the mean signal of the X-ray, and $\sigma_n$ represents the pixel noise arising from fluctuations of the baseline and the threshold.
$B$ is an offset to account for a high-energy contamination of the measured spectrum from primary X-rays Compton-scattered at the target, which are observed for the low energy X-rays from titanium.
For higher energies, it is set to zero.
At very low thresholds, the hit count drops significantly due to an over-saturation of the readout caused by a strongly rising noise rate.
Consequently, this region is excluded from the fit of the s-curve.

The resulting $\mu$ and $\sigma_n$ from the fit function are filled into histograms as shown in Figures~\ref{fig:mu1D} and \ref{fig:sigma1D}, which contain one entry from the fits to the s-curve of each pixel in the matrix.
The histograms show normal distributions, which are fitted with a Gaussian to obtain the mean and the spread of each distribution: $\overline{\mu}\pm\Delta \overline{\mu}$ and $\overline{\sigma_n}\pm\Delta \overline{\sigma_n}$.
The non-gaussian tails visible on the left in Figure~\ref{fig:mu1D} and on the right in Figure~\ref{fig:sigma1D} originate from noisy pixels.

\subsection{Gain and Baseline}
\begin{figure}[t]
\begin{center}
\begin{subfigure}[b]{0.3\textwidth}
\includegraphics[width=\textwidth]{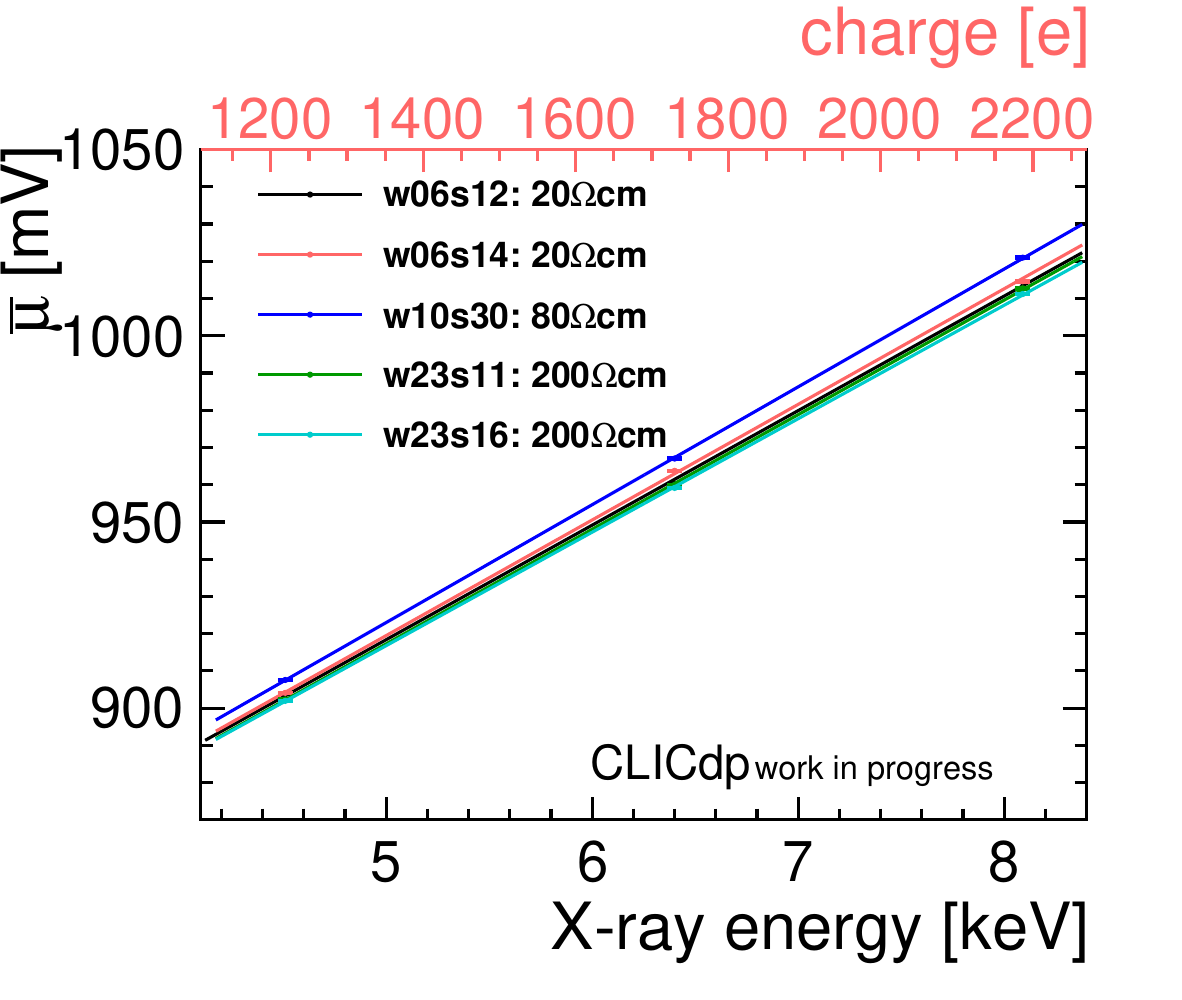}
\caption{\label{fig:signal-vs-energy}Signal size vs.~X-ray energy.}
\end{subfigure}\hfill%
\begin{subfigure}[b]{0.3\textwidth}
\includegraphics[width=\textwidth]{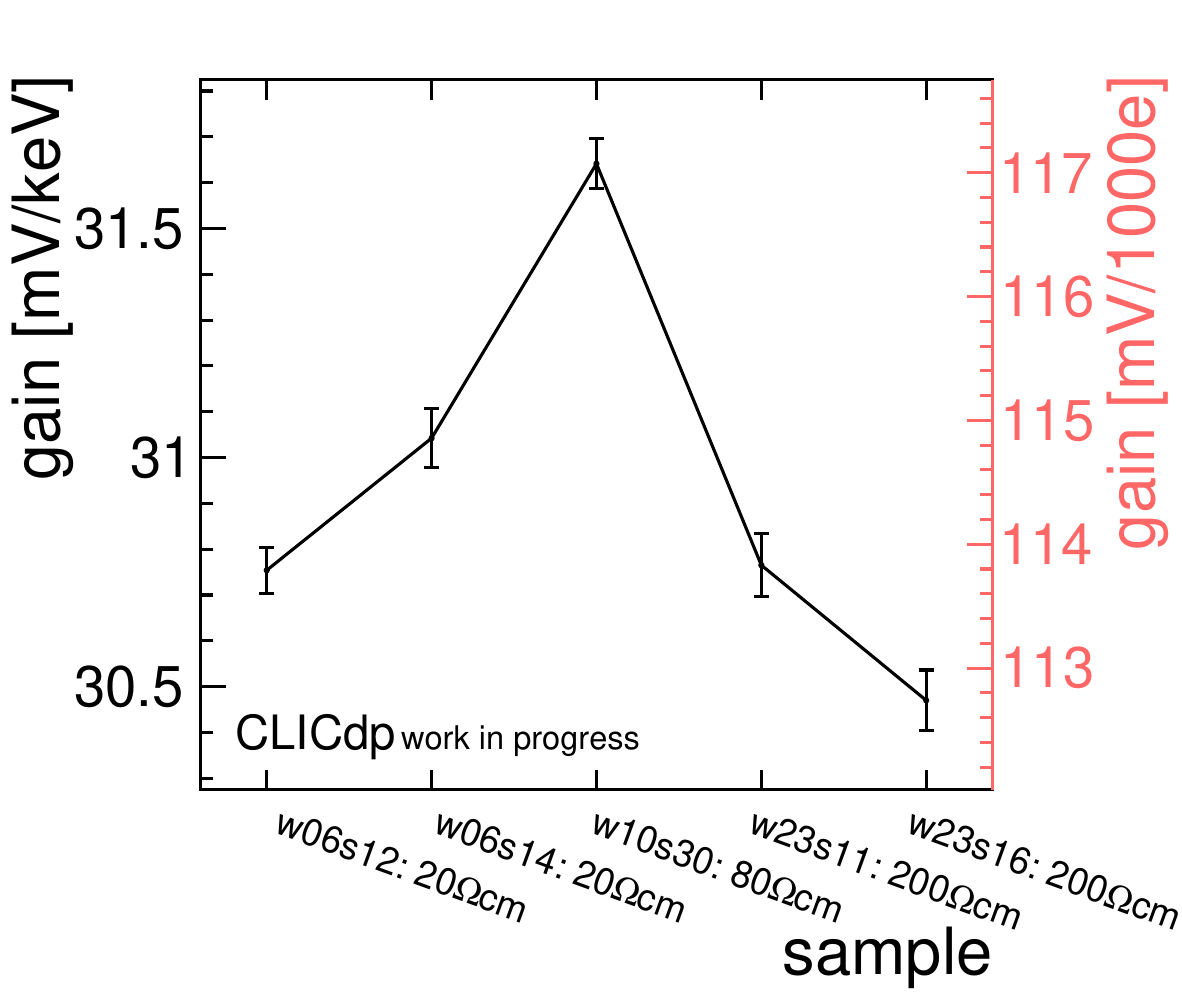}
\caption{\label{fig:gain}Gain.}
\end{subfigure}\hfill%
\begin{subfigure}[b]{0.3\textwidth}
\includegraphics[width=\textwidth]{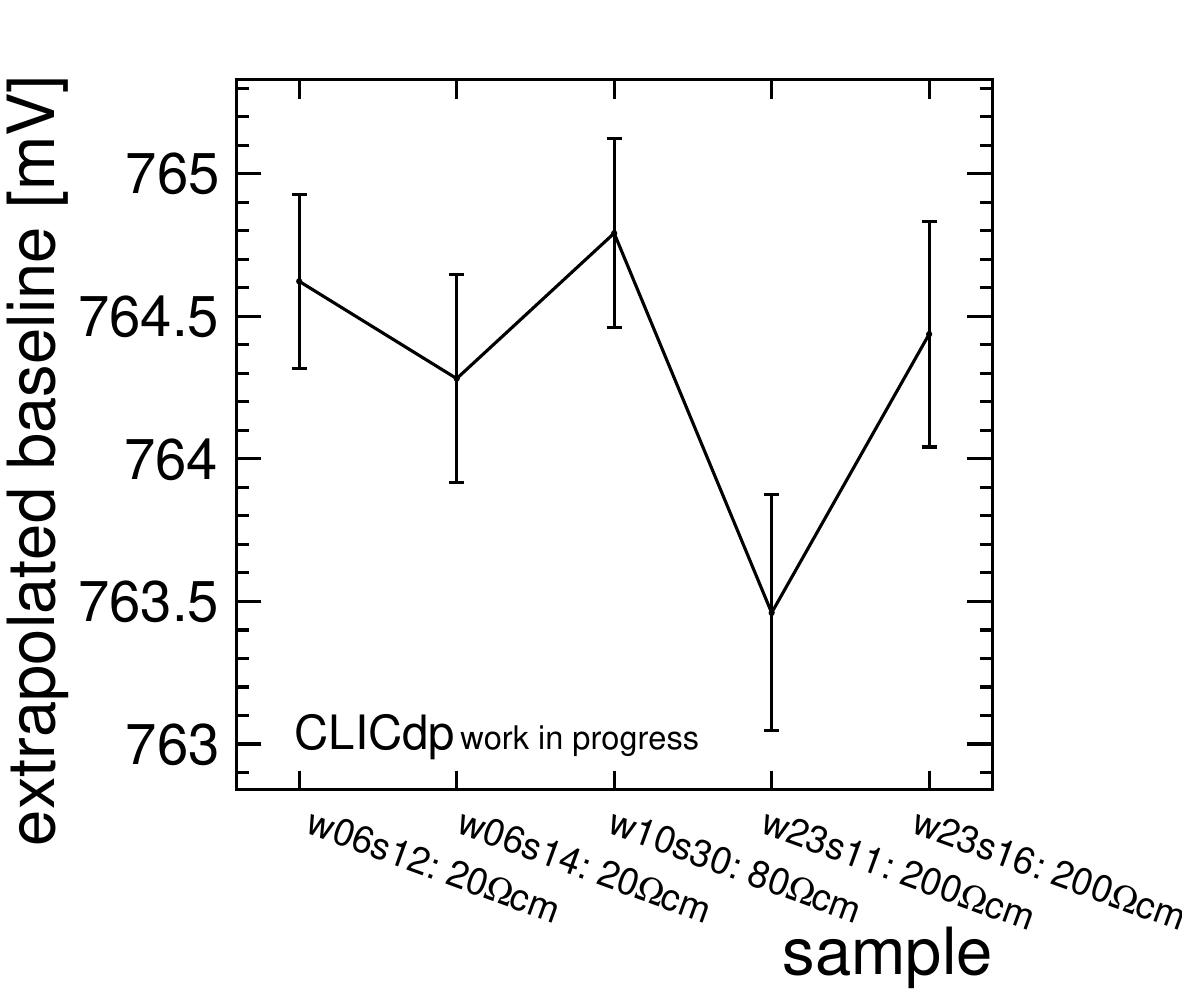}
\caption{\label{fig:baseline}Extrapolated baseline.}
\end{subfigure}
\caption{Signal vs.~X-ray energy with a linear fit and the resulting gain (slope) and extrapolated baseline ($y$-intercept) for different samples and various fluorescent X-ray energies at a bias voltage of \SI{-50}{V}. The error bars correspond to the statistical uncertainties on the fit parameters.}
%\vspace{-0.5cm}
\end{center}
\end{figure}

Since soft X-rays are absorbed completely, the amount of deposited energy $E$ is well-defined and can be converted into the signal charge $Q$ corresponding to the number of created electron-hole pairs.
An average energy of \SI{3.7}{eV} is required to generate one electron-hole pair~\cite{fraser-electron-hole-pair-energy}.
Figure~\ref{fig:signal-vs-energy} shows the $\overline{\mu}$ values obtained for the different X-ray targets.
A first-order polynomial is fitted for all samples:
% \vspace{-0.5cm}
\begin{equation}
\label{eq:thres-calibration}
\overline{\mu} = g~\text{\footnotesize{[mV/keV]}} \cdot E + b = g~\text{\footnotesize{[mV/1000e]}} \cdot Q + b
\end{equation}
where $b$ denotes the extrapolated baseline, i.e.~the $y$-intercept of the polynomial.
The slope $g$ of the fit function can be interpreted as the signal gain, which is summarised in Figure~\ref{fig:gain} for all presented samples.
% The \SI{80}{\ohm cm} sample has a larger gain than the \SI{20}{\ohm cm} samples, whereas the \SI{200}{\ohm cm} samples show a smaller gain.
% If at all, the charge collection efficiency would be expected to be larger for higher substrate resistivities because of a smaller recombination probability.
Since no clear trend of the gain with the substrate resistivity can be seen, it can be concluded that the observed differences stem from sample-to-sample or wafer-to-wafer variations and are dominated by the electronics.

Figure~\ref{fig:baseline} summarises the extrapolated baseline, i.e.~the $y$-intercept of the linear fit functions for all samples.
It is observed that it differs notably from the externally applied baseline of \SI{800}{mV}.
This effect is consistent with an expected voltage offset within the in-pixel comparator, which can be $\mathcal{O}(\text{few  } \SI{10}{mV})$~\cite{communication-ivan}.
% However, this regime is difficult to access experimentally due to a strongly increasing noise rate at low thresholds causing an over-saturation of the readout.

The inversion of Equation~\ref{eq:thres-calibration} can be used to determine the signal charge for a given threshold:
%\vspace{-0.5cm}
\begin{linenomath*}
\begin{equation}
\label{eq:threshold-conversion-inverted}
Q = \frac{\overline{\mu} - b}{g}
\end{equation}
\end{linenomath*}
with the statistical uncertainty obtained by Gaussian error propagation.

Averaging over all samples a gain of $\overline{g} = \SI{114.5\pm1.5}{mV/1000e}$ and an extrapolated baseline of $\overline{b} = \SI{764.3\pm0.6}{mV}$ have been measured.
It is important to note that the gain strongly depends on the chip configuration, in particular those parameters which regulate the current of the charge-sensitive amplifier in the pixel.

Using the gain as a conversion factor, the threshold dispersion $\overline{\sigma_\mu}$ and the pixel noise $\overline{\sigma_n}$ determined previously can now be translated into an equivalent charge.
The standard deviation $\sigma_\mu~\sim\SI{129}{e^-}$ is the threshold dispersion, i.e.~the variation of the the detection threshold across the pixel matrix.
The mean of $\sigma_n\sim\SI{148}{e^-}$ is the average pixel noise.
For an energy deposition of \SI{6.4}{keV}, this results in a signal-to-noise ratio of
\begin{linenomath*}
\begin{equation}
SNR = \frac{\mu - b}{\sigma_n} \approx 11.0\pm1.2.
\end{equation}
\end{linenomath*}

\vspace{-0.5cm}
\section{Test-beam Performance Measurements}
Performance measurements of the \apx have been carried out at the DESY-II test-beam facility~\cite{desy2} using EUDET-type reference telescopes~\cite{eudet}. For the chosen electron beam momentum of \SI{5.4}{GeV}, these yield a track pointing resolution of \SIrange{2}{5}{\micro m}, depending on the plane spacing, which allows to study in-pixel effects.
With an additional Timepix3 plane~\cite{timepix3}, a track time resolution of \SI{1.1}{ns} is achieved.
The \apx was controlled and read out with the Caribou system~\cite{caribou}, and the reconstruction and analysis was carried out using the Corryvreckan framework~\cite{corry-publication}.

\subsection{Spatial and Time Resolution and Hit Detection Efficiency}
Previous studies~\cite{proceedings-instr20} have shown that the \apx reaches a binary spatial resolution limited by its pixel pitch and a timing resolution down to \SI{6.7}{ns} for the \SI{200}{\ohm cm} samples after a row-dependent delay and a timewalk correction.
Lower substrate resistivities lead to a slower timing.
Samples with all substrate resistivities can be operated at a high detection efficiency above \SI{99}{\percent}, whereas the efficiency at high thresholds and low bias voltages remains larger for higher substrate resistivities.

\begin{figure}[b]
\begin{center}
\begin{minipage}[b]{0.55\textwidth}
\centering
\includegraphics[width=\textwidth]{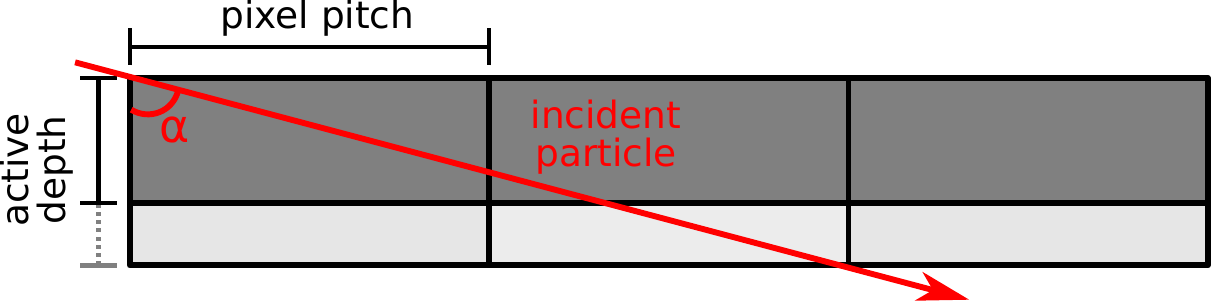}
\vspace{0.25cm}
\caption{\label{fig:rotation-model}Schematic drawing of the track incidence angle dependence on the expected cluster size.\hfill\newline\newline}
\end{minipage}\hfill%
\begin{minipage}[b]{0.40\textwidth}
\centering
\includegraphics[width=0.9\textwidth]{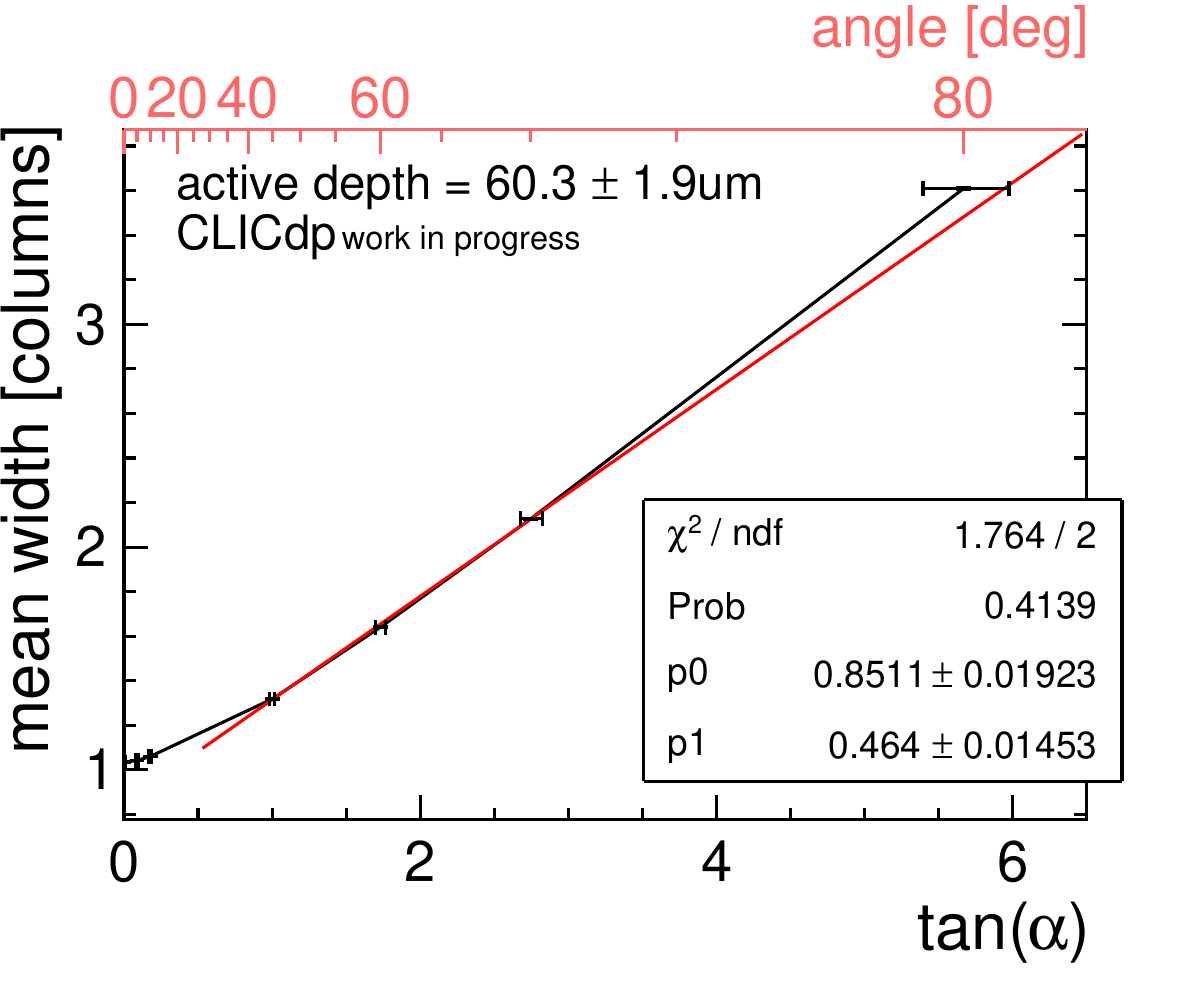}
\vspace{-0.5cm}
\caption{\label{fig:clustersize-vs-angle}Angle dependence of the mean cluster size in column direction with a linear fit to extract the depletion depth.}
\end{minipage}
\end{center}
\end{figure}

\subsection{Active Depth Determination}
Inclined tracks are expected to lead to increased cluster sizes because a particle penetrates several adjacent pixels while passing through the detector material.
As illustrated in Figure~\ref{fig:rotation-model}, this is dependent on the incidence angle $\alpha$ as well as the pixel pitch and the active depth $d_\text{active}$.
Here, the active depth refers to the depleted volume plus a possible additional layer below the depletion depth, from which charge may be collected  by diffusion into the depletion region.
In this simple geometrical model, the average cluster width in column/row direction is given by
\begin{linenomath*}
\begin{equation}
\label{eq:clustersize-angle}
\text{cluster width}_\text{column/row} = \frac{d_\text{active}}{\text{pitch}_\text{column/row}} \cdot \tan(\alpha).
\end{equation}
\end{linenomath*}
In turn, a measurement of the angle dependence of the cluster size can be used to obtain an estimation of the active depth.
This represents a simplified model, which neglects possible sub-threshold effects as well as lateral diffusion.
This is an appropriate approximation for the \apx, for which the mean cluster size is only marginally larger than one~\cite{phd_kroeger}.

Figure~\ref{fig:clustersize-vs-angle} shows the mean cluster width in column direction plotted against the tangent of the rotation angle.
Using equation~\ref{eq:clustersize-angle}, the active depth can be retrieved from the slope of a linear fit by dividing through the pixel pitch in the respective dimension.
This yields an estimation of $\SI{60.3\pm1.9}{\micro m}$ at a bias voltage of \SI{-75}{V}.
The comparison with TCAD studies~\cite{phd_meneses} suggests that the substrate resistivity lies around \SIrange{300}{400}{\ohm cm} compared to the nominal value of \SI{200}{\ohm cm}.
A possible range of \SIrange{100}{400}{\ohm cm} is stated by the manufacturer due to deviations of the production parameters from the standard process~\cite{atlaspix-manual}.

% \begin{figure}[h]
% \begin{center}
% \begin{minipage}[b]{0.38\textwidth}
% \centering
% \includegraphics[width=\textwidth]{rotation_sketch}
% \caption{\label{fig:rotation-model}Schematic drawing of the track incidence angle dependence on the expected cluster size.\hfill\newline}
% \end{minipage}\hfill%
% \begin{minipage}[b]{0.33\textwidth}
% \centering
% \includegraphics[width=\textwidth]{DESY_DepletionDepthCol}
% \end{minipage}
% \begin{minipage}[b]{0.225\textwidth}
% \centering
% \caption{\label{fig:clustersize-vs-angle}Angle dependence of the mean cluster size in column direction with a linear fit to extract the depletion depth.}
% \end{minipage}\hfill%
% \end{center}
% \end{figure}

\section{Conclusions}
It was shown that an X-ray based calibration is crucial for the the conversion of the detection threshold and noise from an applied voltage to equivalent charge.
A higher substrate resistivity yields a higher efficiency and a better time resolution.
The determination of the active depth implies that the substrate resistivity exceeds \SI{200}{\ohm cm} in accordance with TCAD simulations~\cite{phd_meneses}.

The obtained results confirm that HV-MAPS is a suitable technology for future tracking detectors.
It is now also investigated as a candidate technology for other experiments such as the MightyTracker Project for the LHCb Upgrades Ib and II~\cite{mighty-tracker}.

\ack{
The measurements leading to these results have been performed at the Test Beam Facility at DESY Hamburg (Germany), a member of the Helmholtz Association (HGF).\\
This work has been sponsored by the Wolfgang Gentner Programme of the German Federal Ministry of Education and Research (grant no. 05E15CHA and 05E18CHA).
}

\section*{References}
\bibliography{references}

\end{document}